\documentclass[12pt]{iopart}

\usepackage{iopams}
\usepackage{cite}
\usepackage{graphicx}
\usepackage{subfigure}

\newcommand{\aof}{\mathring{a}_{of}^{(3)}}
\begin{document}
	\title{Lorentz violation effects in $2\nu\beta\beta$ decay}
	\author{O. Ni\c{t}escu$^{1,2,3}$\footnote{Authors contributed equally to this work.} , S. Ghinescu$^{1,2,3}$\footnotemark[\value{footnote}] and
		S. Stoica$^{1,2}$}
	\address{$^1$ International Centre for Advanced Training and Research in Physics, P.O. Box MG12, 077125-Magurele, Romania.}
	\address{$^2$Horia Hulubei National Institute of Physics and Nuclear Engineering, P.O. Box MG6, 077125-Magurele, Romania}
	\address{$^3$University of Bucharest, Faculty of Physics, P.O. Box MG11, 077125-Magurele, Romania}
	\eads{\mailto{sabin.stoica@cifra.infim.ro}}

 \begin{abstract}
	Observable effects for the Lorentz invariance violation (LIV) at a low energy scale can also be investigated in double beta decay (DBD). For example, by comparing the theoretical predictions with a precise analysis of the summed energy spectra of electrons in $2\nu\beta\beta$ decay, one can constrain the $\aof$ coefficient that governs the time-like component of the Lorentz invariance violating operator that appears in the Standard Model extension theory.
	In this work, we perform calculations of the phase space factors and summed energy spectra of electrons as well as of their deviations due to LIV necessary in such experimental investigations. The Fermi functions needed in the calculation are built up with exact electron wave functions obtained by numerically solving the Dirac equation in a realistic Coulomb-type potential with the inclusion of the finite nuclear size and screening effects. We compared our results with those used in previous LIV investigations that were obtained with approximate (analytical) Fermi functions and found differences of up to $30\%$ for heavier nuclei. Our work includes eight experimentally interesting nuclei. Next, we estimate and discuss the uncertainties of our calculations associated with uncertainties in Q-values measurements and the differences raised from the inclusion of the kinematic terms in the formalism. Finally, we provide the ratio between the standard phase space factors and their LIV deviations and the energies where the LIV effects are expected to be maximal. We expect our study to be useful in the current LIV investigations in $2\nu\beta\beta$ decay and to lead to improved constraints on the $\aof$ coefficient. 
	
\end{abstract} 

\submitto{\jpg}
\maketitle

\section{Introduction}
There is a currently increasing interest in testing possible violations of fundamental symmetries in physical processes, one of them being the Lorentz invariance. The general theory that incorporates LIV is the Standard-Model extension (SME), which was developed by including Lorentz-violating terms in the Lagrange density \cite{CK-PRD55,CK-PRD58,KOS-PRD69,KR-PMP2011}. These terms are constructed as coordinate-invariant products of a Lorentz invariance violating operator and a coefficient that controls the LIV size. The LIV operators can be of arbitrarily large dimension. Still, a particular interest represents the minimal SME, a limiting case of the SME theory that includes only LIV operators of mass dimension four or less \cite{KR-PMP2011}.
Direct observation of LIV implies the investigation of physics at the Plank scale, which is not possible at the moment. However, it may be possible that LIV effects can manifest at a low energy scale and be potentially observable with current or near-future experimental techniques. For example, the experiments dedicated to the investigation of neutrino properties can be a framework where visible effects of LIV can be searched. The operators that couple to neutrinos in SME affect neutrino oscillations, neutrinos velocity and the electron spectra in beta and double-beta decays \cite{KM-PRD69,Adam2012,Diaz-PRD88,Diaz-AHEP,Diaz-PRD89}. Effects of LIV have been searched in many neutrino oscillation experiments as Double-Chooz \cite{DC-PRD86}, MiniBooNE \cite{MBoone-PLB718}, IceCube \cite{IC-PRD82}, MINOS \cite{Minos-PRD85}, SuperKamiokande \cite{SK-PRD91} which obtained constraints on the corresponding coefficients. 
However, LIV in the neutrino sector can also be induced by a Lorentz-violating operator in SME, called countershaded operator, which does not affect the neutrino oscillations and which also breaks the CPT symmetry. The LIV effects related to this operator are controlled by the oscillation-free (of) coefficient with four components, one time-like $(a^{(3)}_{of})_{00}$ and three space-like $(a^{(3)}_{of})_{1m}$, with $m=0, \pm 1$ \cite{KM-PRD69}. Non-zero values of $(a^{(3)}_{of})_{00}$ coefficient would produce small deviations in the shape of the energy spectra of electrons emitted in beta and double-beta decays \cite{Diaz-PRD89}. Such investigations were recently performed in the DBD collaborations EXO \cite{EXO-200-PRD93}, GERDA \cite{GERDA-PhDThesis}, SuperNEMO \cite{NEMO-3-2019}, CUORE \cite{CUORE-2019,CUORE-PhDThesis} and CUPID-0\cite{CUPID-0-PRD100}, and the non-observation of LIV effects resulted in constrains of the isotropic coefficient $\aof=(a^{(3)}_{of})_{00}/\sqrt{4\pi}$ \cite{Diaz-PRD88}. However, in most of these analyses, the necessary theoretical ingredients were calculated with approximate (analytical) Fermi functions \cite{Primakoff-1959,Haxton-1984,Doi-1985,Suhonen-1998}.  

In this paper, we perform calculations of the phase space factors (PSF), summed energy spectra of electrons and their deviations necessary in LIV investigations from DBD experiments. The Fermi functions needed in the calculation are built with exact electron wave functions obtained by numerically solving a Dirac equation in a realistic Coulomb-type potential, with the inclusion of finite nuclear size and screening effects using a method described in our previous papers\cite{SM-2013,MPS-2015}. We compared our results with those used in previous LIV investigations, which were obtained with approximate (analytical) Fermi functions, and found differences up to $30\%$ for heavier nuclei. Our study includes the nuclei $^{48}$Ca, $^{76}$Ge, $^{82}$Se, $^{100}$Mo, $^{110}$Pd, $^{116}$Cd, $^{130}$Te and $^{136}$Xe. Next, we estimate and discuss the uncertainties in our calculations associated with uncertainties in Q-values measurements and the differences raised from the inclusion of the kinematic terms in the formalism. 
The Q-values and their uncertainties were obtained for each nucleus by averaging values reported in the literature by using a statistical procedure recommended by the Particle Data Group \cite{PDG-statistics}. Finally, we provide for each nucleus the ratio between the standard PSF and their LIV deviations and the energies where the LIV effects are expected to be maximal. We expect our study to be useful in the current LIV investigations in $2\nu\beta\beta$ decays and to lead to improved constraints on the $\aof$ coefficient.

\section{Formalism}   
The interactions of neutrinos with the  countershaded operator modify their four-component momentum from the standard expression $q^{\alpha} = (\omega, {\bf q})$  to  $q^{\alpha} = (\omega, {\bf q} + {\bf a}^{(3)}_{of}-\mathring{a}^{(3)}_{of} \bf \hat{q}) $ \cite{KR-PMP2011,Diaz-PRD89,KT-PRL102}. Considering this modification, the $2\nu\beta\beta$ decay rate can be expressed as a sum of two components:  
\begin{equation} 
\label{eq:DecayRateTotal}
\Gamma = \Gamma_0 + \delta \Gamma,
\end{equation}
where $\Gamma_0$ is the standard decay rate and $\delta \Gamma$ is the perturbation induced by LIV.
As known, the standard $2\nu\beta\beta$ decay rate can be expressed to a good approximation as follows \cite{VES-2012}:
\begin{equation} 
\label{eq:StandardDecay}    \Gamma_0 = g^4_A|m_ec^2M^{2\nu}|^2G^{2\nu}   \\ 
\end{equation}
In this expression $g_A$ is the axial vector constant, $M^{2\nu}$ (in MeV$^{-1}$) is the nuclear matrix element and $G^{2\nu}$ is the PSF of the transition. Lorentz-violating effects in DBD appear as kinematical effects modifying only the PSF. Thus, the decay rate induced by LIV can be expressed in the form: 
\begin{equation}
\label{eq:LVDecay}   \delta \Gamma = g^4_A|m_ec^2M^{2\nu}|^2\delta G^{2\nu} 
\end{equation}
where $\delta G^{2\nu}$ is the PSF perturbation due to LIV. In what follows, we adopt the natural units ($\hbar=c=1$), and we write the energy variables in units of $m_e$. For a $2\nu\beta\beta$ transition to the ground state (g.s.) of the daughter nucleus, the PSF expression reads:\\
\begin{eqnarray}
\label{eq:G2nuVS}
\eqalign{G^{2\nu}&=C_1
	\int_{0}^{Q} d\epsilon_1 \int_{0}^{Q-\epsilon_1} d\epsilon_2 \int_{0}^{Q-\epsilon_1-\epsilon_2} 
	d\omega_1 
	\\
	&  \times F(Z_{f},\epsilon_1)F(Z_{f},\epsilon_{2})
	\sqrt{\epsilon_1 (\epsilon_1 + 2)} (\epsilon_1 + 1)\sqrt{\epsilon_2 (\epsilon_2 + 2)} (\epsilon_2 + 1) \\
	&\times \omega_1^2(Q-\epsilon_1-\epsilon_2-\omega_1)^2\left(\langle K_N\rangle^2+\langle L_N\rangle^2+
	\langle K_N\rangle \langle L_N\rangle\right)}
\end{eqnarray}
\noindent
where $C_1= (\tilde{A}^2 G_F^4 |V_{ud}|^4 m_e^9 )/(96\pi^7\rm{ln}2)$, $G_F$ is the Fermi coupling constant and $V_{ud}$ the first element of the CKM matrix. $\epsilon_{1,2}$ and $\omega_1$ are the energies of the electrons and of one antineutrino emitted in the decay and $F(Z, \epsilon)$ is the Fermi function.
$\langle K_N\rangle$, $\langle L_N\rangle$ are kinematic factors that depend on the electrons 
$(\epsilon_{1,2})$ and antineutrinos ($\omega_{1,2}$) energies, on the g.s. energy $E_I$ of 
the parent nucleus and on an averaged energy $\langle E_N \rangle$ of the excited states in the intermediate nucleus (closure approximation) \cite{Haxton-1984}. 
\begin{equation}
\label{eq:KnDef}
\langle K_N\rangle=
{1\over \epsilon_1+\omega_1+\langle E_N\rangle-E_I+1}+
{1\over \epsilon_2+\omega_2+\langle E_N\rangle-E_I+1}
\end{equation}
\begin{equation}
\label{eq:LnDef}
\langle L_N\rangle=
{1\over \epsilon_1+\omega_2+\langle E_N\rangle-E_I+1}+
{1\over \epsilon_2+\omega_1+\langle E_N\rangle-E_I+1}
\end{equation}
Here, the difference in energy in the denominator can be obtained from the approximation
$\tilde{A}^2=[Q/2+\langle E_N\rangle -E_I+1]^2$, where
$\tilde{A}=1.12A^{1/2}$ (in MeV) gives the energy of the giant Gamow-Teller resonance in the intermediate nucleus.

In order to evaluate $\delta G^{2\nu}$ from Eq. (4), we make use of the  differential element of the antineutrino momentum which, from $d^{3}q = 4\pi\omega^{2}d\omega$ in the standard case, now becomes $d^{3}q = 4\pi(\omega^{2}+2\mathring{a}^{(3)}_{of}\omega)d\omega$. Thus LIV contribution reads:

\begin{eqnarray}
\label{eq:DeltaG2nuVS}
\eqalign{\delta G^{2\nu}&=10 \mathring{a}^{(3)}_{of}C_2
	\int_{0}^{Q} d\epsilon_1 \int_{0}^{Q-\epsilon_1} d\epsilon_2 \int_{0}^{Q-\epsilon_1-\epsilon_2} 
	d\omega_1 
	\cr
	&  \times F(Z_{f},\epsilon_1)F(Z_{f},\epsilon_{2})
	\sqrt{\epsilon_1 (\epsilon_1 + 2)} (\epsilon_1 + 1)\sqrt{\epsilon_2 (\epsilon_2 + 2)} (\epsilon_2 + 1) \cr
	& \times	\omega_1(Q-\epsilon_1-\epsilon_2-\omega_1)^2\left(\langle K_N\rangle^2+\langle L_N\rangle^2+
	\langle K_N\rangle \langle L_N\rangle\right)} 
\end{eqnarray}\\
where $C_2= (\tilde{A}^2 G_F^4 |V_{ud}|^4 m_e^8 )/(240\pi^7\rm{ln}2)$.
The isotropic coefficient $\mathring{a}^{(3)}_{of}$, which governs the Lorentz violation strength, is the quantity of interest.

We note here that by making the approximation: 
\begin{equation}
\label{eq:KnLnApprox}
\langle K_N\rangle\simeq \langle L_N\rangle\simeq\frac{2}{E_I-\langle E_N\rangle-\left(Q/2+1\right)}
\end{equation}
together with the approximation on $\tilde{A}$ and integrating over the energy of the antineutrino $\omega_1$, one retrieves simplified expressions for the PSF which were used in many previous DBD works (see for example \cite{Suhonen-1998} and references therein) and also in the previous LIV analyzes \cite{EXO-200-PRD93,CUPID-0-PRD100,NEMO-3-2019}:\\

\begin{eqnarray}
\label{eq:PhaseSpaceIntegral_2nu}
\eqalign{ G^{2\nu} & = C_3 \int _0^{Q} d\epsilon_1  \int _0 ^{Q - \epsilon_1} d\epsilon_2  F(Z_{f},\epsilon_1)F(Z_{f},\epsilon_{2}) \cr
	& \times  \sqrt{\epsilon_1 (\epsilon_1 + 2)} (\epsilon_1 + 1)\sqrt{\epsilon_2 (\epsilon_2 + 2)} (\epsilon_2 + 1) (Q - \epsilon_1 - \epsilon_2)^5}
\end{eqnarray}

\begin{eqnarray}
\label{eq:PSF_LV}
\eqalign{ \delta G^{2\nu}& = 10 \mathring{a}^{(3)}_{of} C_4 \int _0^{Q} d\epsilon_1 \int _0 ^{Q - \epsilon_1} d\epsilon_2 F(Z_{f},\epsilon_1)F(Z_{f},\epsilon_{2}) \cr
	& \times \sqrt{\epsilon_1 (\epsilon_1 + 2)} (\epsilon_1 + 1) \sqrt{\epsilon_2 (\epsilon_2 + 2)} (\epsilon_2 + 1) (Q - \epsilon_1 - \epsilon_2)^4} \label{PhaseSpaceIntegral}
\end{eqnarray} 
where $C_3 = (G_F^4 |V_{ud}|^4 m_e^9 )/(240\pi^7\rm{ln}2)$ and $C_4 = (G_F^4 |V_{ud}|^4 m_e^8 )/(240\pi^7\rm{ln}2)$.   

As seen from the above expressions, the main ingredients in the PSF calculation are the Fermi functions, which encodes the distortion of the electron wave functions due to the Coulomb field of the daughter nucleus, the kinematic factors $\langle K_N\rangle$, $\langle L_N\rangle$ and the Q-values. 
In the next sections, we refer to their use in the computation of the quantities relevant for analysing of LIV effects in $2\nu\beta\beta$ decay.

\section{Fermi functions}
\label{sec:Fermi}

In this section, different approximation schemes used for the calculation of the Fermi function are briefly presented.
\subsection{The approximation scheme A}
In an early derivation of the $2\nu\beta\beta$ decay rate Primakoff\& 
Rosen \cite{Primakoff-1959} and Konopinski\cite{Konop-1966}, 
considered a non-relativistic movement of the emitted electrons in a 
Coulomb potential given by a point-like nucleus. The correction to the 
electron wave function can be derived by taking the square of the 
ratio between the Schr\"odinger scattering solution for a point charge 
$Z_f$ and a plane wave, evaluated at the origin. In this case,
\begin{equation}
F(Z_{f},\epsilon) = F^{NR}(Z_f,\epsilon)=\frac{2\pi\eta}{1-e^{-2\pi\eta}}
\end{equation}
where $\eta=\pm\alpha Z_f \epsilon/p$ ("+" corresponds to electrons and "-" to positrons), $\alpha\simeq1/137$ is the fine structure constant, $\epsilon$ is the energy of the outgoing particle and $p=\left|\vec{p}\right|$ is their momentum.

Although this approximation fails badly for heavy nuclei (for example, a factor of 5 enhancement in the $^{130}$Te phase space factor \cite{Haxton-1982}), its use has the advantage that DBD rate can be integrated analytically and the obtained expression of the decay rate is a polynomial in powers of the Q-value.  
\subsection{The approximation scheme B}  
Adopting a relativistic treatment, the Fermi function is obtained as solution of a Dirac equation in a Coulomb potential given by a point charge (the finite size of the nucleus is neglected). It can still be expressed in an analytical form as:  
\begin{eqnarray}\label{eq:relativisticFermi}
\label{eq:FermiFunc}
F(Z_{f},\epsilon) = F_0(Z_f,\epsilon) =  4(2pR_A)^{2(\gamma-1)}e^{\pi\eta}\frac{\left|\Gamma(\gamma+\rm i \eta)\right|^{2}}{\left[\Gamma(2\gamma+1)\right]^{2}},
\end{eqnarray}	
with 
\begin{equation}
\gamma=\sqrt{1-(\alpha Z_f)^2},\qquad \eta=\pm\alpha Z_f \epsilon/p
\end{equation}
with the same sign convention as above. $R_A$ is the cut-off radius in the
evaluation of the Dirac equation which is taken to match the radius of
the daughter nucleus (i.e. $R_A=1.2A^{1/3}$ fm). This approximation
scheme was used extensively in the past \cite{Haxton-1984,Doi-1985,Suhonen-1998} and
will also be used here for comparison. We also note that in those works, the Gamma 
functions were computed using approximate expressions. In our case,
the above expression of the Fermi function was computed using 
an exact form of the Gamma function and the results agree (within 
a few percentages) with the ones from Refs. \cite{Haxton-1984,Doi-1985,Suhonen-1998} .

\subsection{The approximation scheme C}
\label{sec:ourmethod}

The Fermi function is built up from the radial solutions of the Dirac equation:   
\begin{equation}
F(Z_{f},\epsilon) = \frac{f_{1}^{2}(\epsilon,R_A)+g^{2}_{-1}(\epsilon, R_A)}{2p^2},
\end{equation}
where $f_1$ and $g_{-1}$ are the radial wave functions of an electron
in the $s_{1/2}$ state evaluated at the nuclear radius $R_A =
1.2 A^{1/3}$, which satisfy the radial Dirac equations \cite{Rose-1961}

\begin{eqnarray}
\label{dirac}\eqalign{ \left(\frac{d}{dr} +
	\frac{\kappa+1}{r}\right)g_{\kappa}(\epsilon,r) =
	(\epsilon-V(r)+1)f_{\kappa}(\epsilon,r) \cr
	\left(\frac{d}{dr} -\frac{\kappa-1}{r}\right)f_{\kappa}(\epsilon,r)=
	-(\epsilon-V(r)-1)g_{\kappa}(\epsilon,r)}
\end{eqnarray}
The relativistic quantum number $\kappa$, takes positive and negative integer values. Total angular momentum of the electron is given by $j_{\kappa}=\left|\kappa\right|-1/2$. The input central potential $V(r)$ includes the finite nuclear size correction and is built from a realistic proton charge density of the daughter nucleus:

\begin{equation}\label{dens}
\rho_e(\vec{r})=\sum_{i}^{}(2j_i+1)v^2_i\left|\psi_i(\vec{r})\right|^2,
\end{equation}
where $j_i$ is the proton spin, $v_i$ is the occupation amplitude of the proton wave function $\psi_i$ of the spherical single particle state $i$, numerically determined as solution of the Schr\"odinger equation with a Wood-Saxon potential. The daughter nucleus potential is given by

\begin{equation}\label{potential}
V(Z,r)=\alpha\hbar c \int
\frac{\rho_e(\vec{r'})}{|\vec{r}-\vec{r'}|}d\vec{r'}.
\end{equation}
This form of the potential includes diffuse nuclear surface corrections. Further,  the screening effect of atomic electrons is taken into account following a prescription described in ref. \cite{Kotila-2012}, by multiplying the expression of $V(r)$ with a function $\phi(r)$, which is solution of the Thomas-Fermi equation
\begin{equation}\label{Thomas-Fermi}
\frac{d^2\phi}{dx^2}=\frac{\phi^{3/2}}{\sqrt{x}},
\end{equation}
with $x=r/b$, $b\approx0.8853a_0Z^{-1/3}$ and $a_0$ is the Bohr radius. The method used to solve numerically the Thomas-Fermi equation is the Majorana method \cite{Esposito-2002}. 
The solutions of the Dirac equation were obtained using our code that was built up following the numerical procedure presented in the refs. \cite{Salvat-1991,Salvat-1995,Salvat-2016}. This treatment of computing the Fermi function was also used by different authors \cite{Kotila-2012,Simkovic-2015,Simkovic-2018}, but the use of a realistic Coulomb potential was introduced only in Refs. \cite{SM-2013,MPS-2015}.  	

\section{Comparison between Fermi functions}

In earlier PSF calculations using approximate (analytical) forms of Fermi functions, the finite nuclear size or screening effects were not considered in computations, which can have a notable impact on the accuracy of the $2\nu\beta\beta$ decay related predictions, required in precise measurements. In this section, we calculate the PSF values and their deviations, using the Equations~(\ref{eq:PhaseSpaceIntegral_2nu}) and (\ref{eq:PSF_LV}) (without the inclusion of kinematic factors, as they were used in previous LIV analyzes). For the LIV perturbation, relevant is the quantity $\delta G^{2\nu}/(10\aof)$, which is independent on the LIV coefficient and is used by the experimentalists to constrain it.

\begin{table}[hb!]
	\lineup
	\caption{\label{tab:PSFAppSchemes}Phase space factors computed with Fermi functions in approximation schemes A, B, C. Values are obtained with Equations~(\ref{eq:PhaseSpaceIntegral_2nu},\ref{eq:PSF_LV}). Q-values are also displayed.}
	\footnotesize\rm
	\begin{tabular}{@{}*{7}{c}}
		\br
		&&\centre{3}{$G^{2\nu}$ in units of $10^{-21}{\rm yr}^{-1}$ }\\
		\ns
		&&\crule{3}\\
		Nucleus&\centre{1}{$Q^{\rm a}(\rm keV)$}&\centre{1}A&\centre{1}{B}&\centre{1}{C}\\
		\mr
		$^{48}$Ca&$4268.070\pm0.076\0$&$12051.21\0\pm2.02\0$&$15941.38\0\pm2.66\0$ &$15212.54\0\pm2.54\0$\\
		$^{76}$Ge&$2039.059\pm0.014\0$&\0\0\0$25.755\pm0.002$&\0\0\0$52.834\pm0.003$&\0\0\0$48.384\pm0.003$\\
		$^{82}$Se&$2997.9\0\0\pm0.3\0\0\0$&\0\0$828.96\0\pm0.75\0$&\0$1779.02\0\pm1.58\0$ &\0$1597.40\0\pm1.42\0$\\
		$^{100}$Mo&$3034.40\0\pm0.17\0\0$&\0$1241.83\0\pm0.62\0$&\0$3822.34\0\pm1.89\0$&\0$3312.62\0\pm1.63\0$\\
		$^{110}$Pd&$2017.85\0\pm0.64\0\0$&\0\0\0$40.32\0\pm0.11\0$&\0\0$163.07\0\pm0.43\0$&\0\0$138.57\0\pm0.37\0$\\
		$^{116}$Cd&$2813.50\0\pm0.13\0\0$&\0\0$776.81\0\pm0.32\0$&\0$3318.62\0\pm1.33\0$&\0$2767.61\0\pm1.11\0$\\
		$^{130}$Te&$2527.515\pm0.260\0$&\0\0$344.24\0\pm0.31\0$&\0$1891.38\0\pm1.66\0$&\0$1538.53\0\pm1.35\0$\\
		$^{136}$Xe&$2458.13\0\pm0.41\0\0$&\0\0$287.56\0\pm0.42\0$&\0$1797.71\0\pm2.55\0$&\0$1441.47\0\pm2.04\0$\\
		\mr
		&&\centre{3}{ $\delta G^{2\nu}/(10\aof)$ in units of $10^{-21}{\rm yr}^{-1}\rm{MeV}^{-1}$}\\
		\ns
		&&\crule{3}\\
		Nucleus&\centre{1}{$Q^{\rm a}(\rm keV)$}&\centre{1}A&\centre{1}B&\centre{1}C\\
		\mr
		
		$^{48}$Ca&$4268.070\pm0.076\0$&$\05318.08\0\pm0.81\0$&$\07002.59\0\pm1.06\0\0$ &$\06674.90\0\pm1.01\0\0$\\
		$^{76}$Ge&$2039.059\pm0.014\0$&\0\0\0$21.754\pm0.001$&\0\0$44.195\pm0.002$&\0\0$40.446\pm0.002$\\
		$^{82}$Se&$2997.9\0\0\pm0.3\0\0\0$&\0\0$497.02\0\pm0.40\0$&$1054.26\0\pm0.84\0$ &\0$945.27\0\pm0.75\0$\\
		$^{100}$Mo&$3034.40\0\pm0.17\0\0$&\0\0$733.70\0\pm0.33\0$&$2219.94\0\pm0.99\0$&$1920.26\0\pm0.85\0$\\
		$^{110}$Pd&$2017.85\0\pm0.64\0\0$&\0\0\0$34.09\0\pm0.08\0$&\0$135.41\0\pm0.32\0$&\0$114.94\0\pm0.27\0$\\
		$^{116}$Cd&$2813.50\0\pm0.13\0\0$&\0\0$489.48\0\pm0.18\0$&$2046.83\0\pm0.74\0$&$1703.76\0\pm0.61\0$\\
		$^{130}$Te&$2527.515\pm0.260\0$&\0\0$238.08\0\pm0.19\0$&$1277.20\0\pm1.00\0$&$1036.90\0\pm0.81\0$\\
		$^{136}$Xe&$2458.13\0\pm0.41\0\0$&\0\0$203.69\0\pm0.27\0$&$1241.58\0\pm1.57\0$&\0$993.53\0\pm1.26\0$\\
		
		\br
	\end{tabular}\\
	$^{\rm a}$ See Appendix for details on Q-values. \\
\end{table}

The results are shown in Table \ref{tab:PSFAppSchemes} wherein the upper part are displayed the PSF values, while in the lower part their deviations. Computations are done for the eight nuclei shown in the first column. The second column indicates the $Q$-values for each decay 
(obtained with the recipe described in the Appendix), while in the next three columns are presented the $G^{2\nu}$ and $\delta G^{2\nu}/(10\aof)$ values 
with Fermi functions obtained in approximation schemes A, B, and C. As expected, there are significant differences between the non-relativistic approximation A and the relativistic ones B, C. The non-relativistic treatment underestimates much the PSF values. However, there are also relevant differences between the PSF values calculated with Fermi functions obtained with B and C treatments. The approximation scheme B overestimates the PSF values as compared with C, and the differences between the two sets of values increase with the nuclear mass from about $8\%$ for $^{48}$Ca to about $30\%$ for $^{136}$Xe. Thus, the non-relativistic treatment of the Fermi functions is inadequate for LIV analyzes. In contrast, the use of exact radial solutions of the Dirac equation for building Fermi functions brings relevant corrections to the $G^{2\nu}$ and $\delta G^{2\nu}$ values as compared with those obtained with analytical expressions.

\begin{figure}[htb!]
	\centering
	\begin{subfigure}
		\centering
		\includegraphics[width=0.49\textwidth]{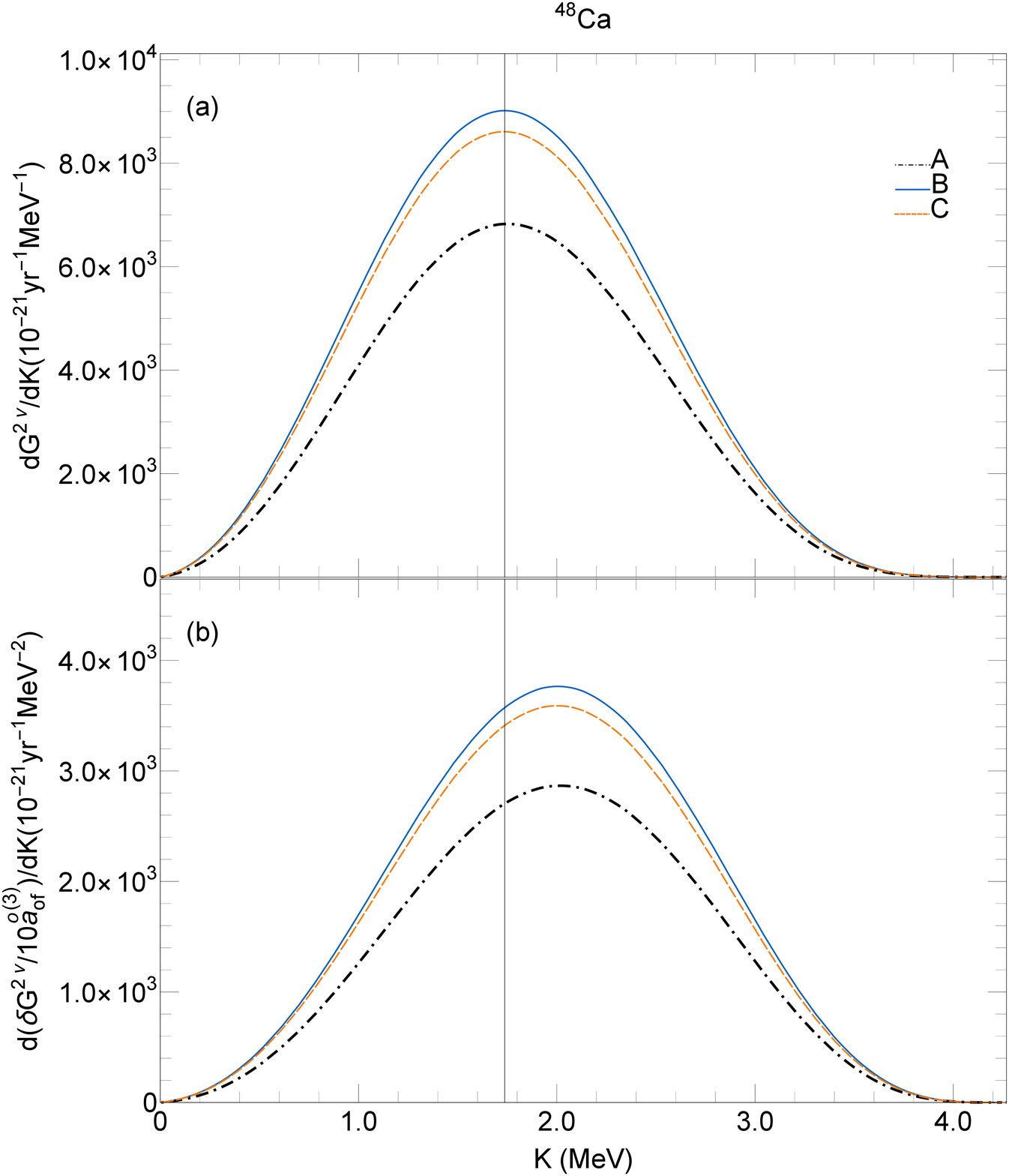}
	\end{subfigure}%
	\begin{subfigure}
		\centering
		\includegraphics[width=0.49\textwidth]{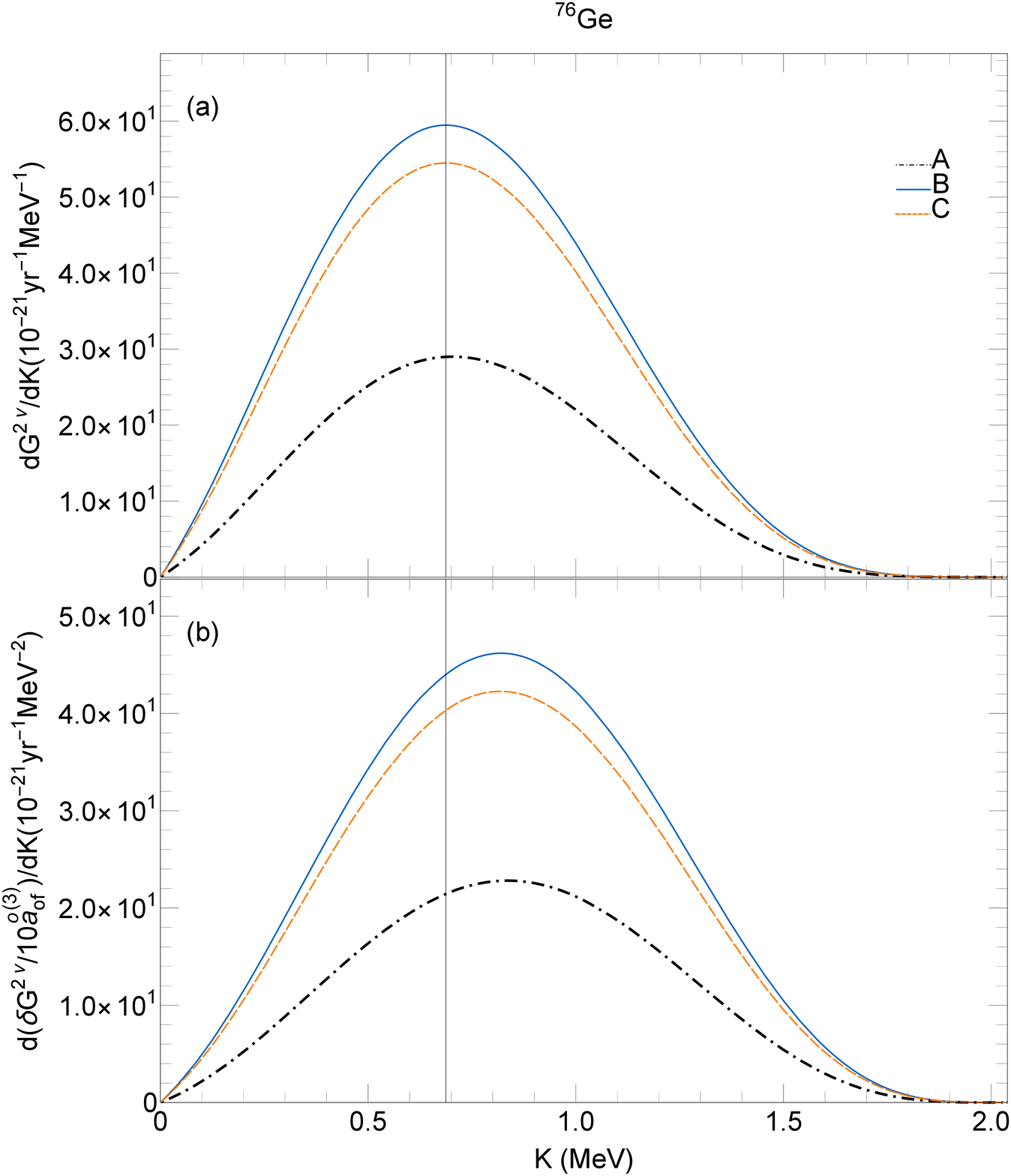}
	\end{subfigure}
	\begin{subfigure}
		\centering
		\includegraphics[width=0.49\textwidth]{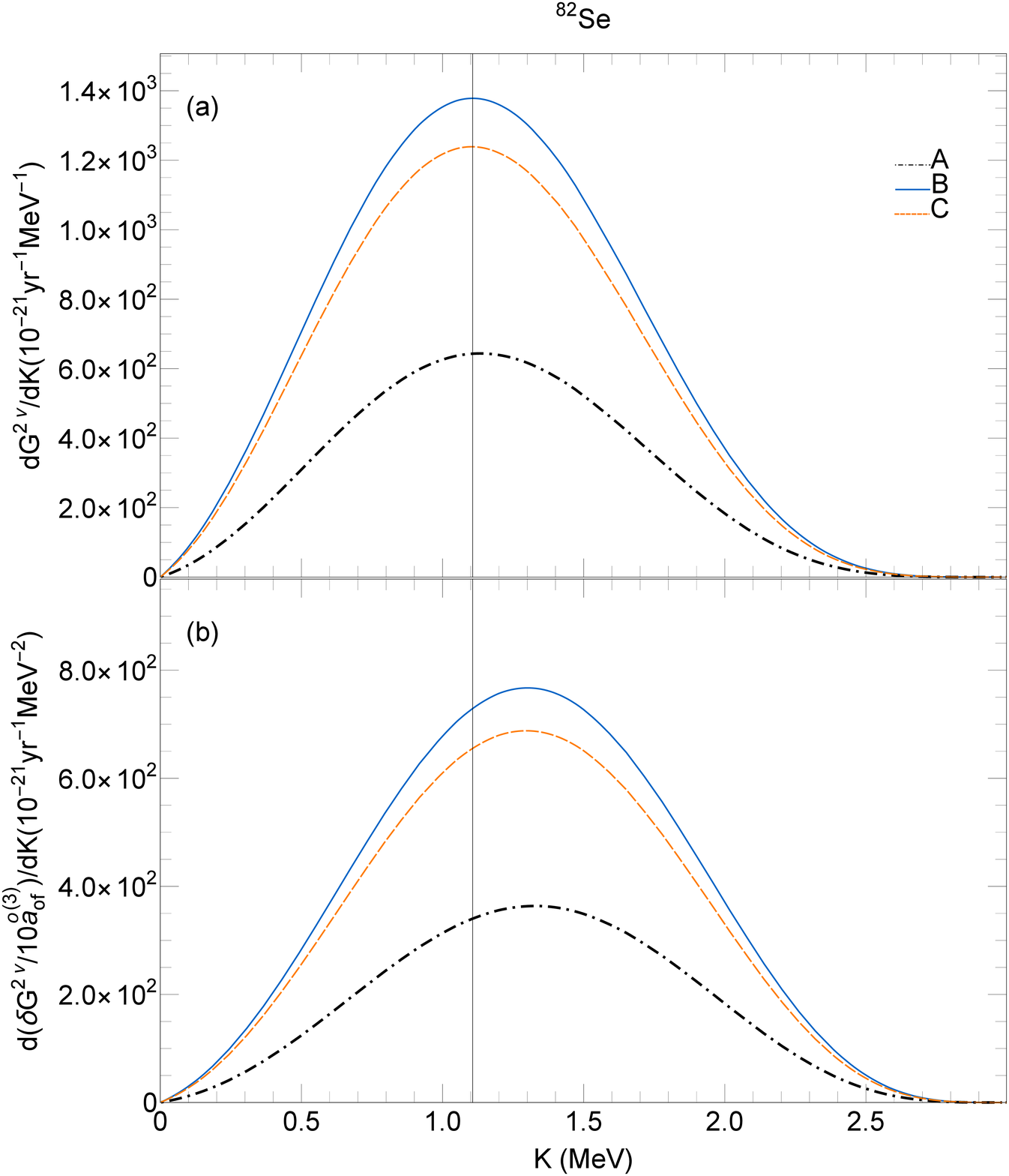}
	\end{subfigure}%
	\begin{subfigure}
		\centering
		\includegraphics[width=0.49\textwidth]{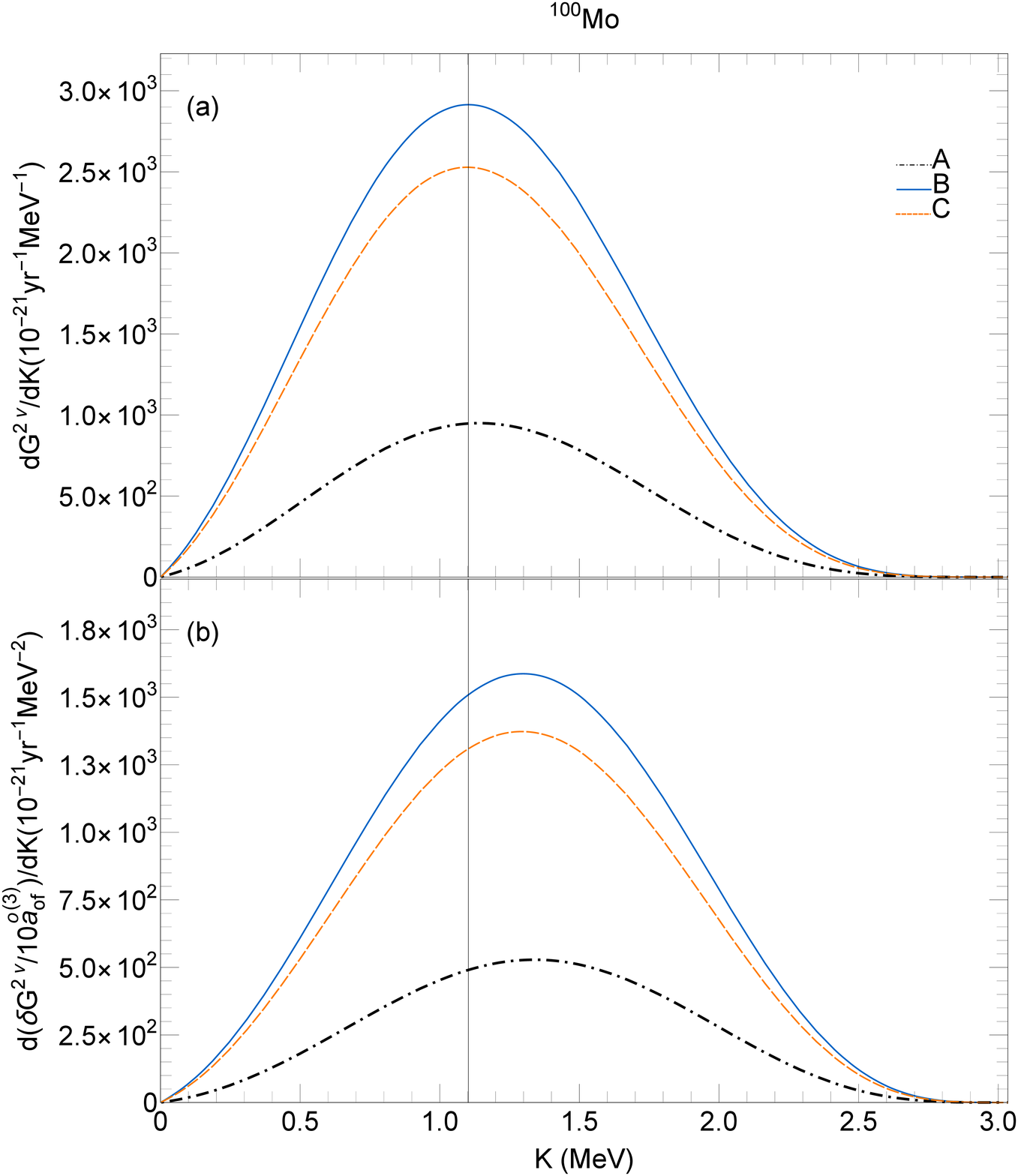}
	\end{subfigure}
	\caption[short]{Summed energy spectra of electrons in the standard $2\nu\beta\beta$ decay (a) and their deviations due to LIV (b) for the nuclei $^{48}$Ca, $^{76}$Ge, $^{82}$Se and $^{100}$Mo.}
	\label{fig:SumSpectra1}
\end{figure}

\begin{figure}[htb!]
	\centering
	\begin{subfigure}
		\centering
		\includegraphics[width=0.49\textwidth]{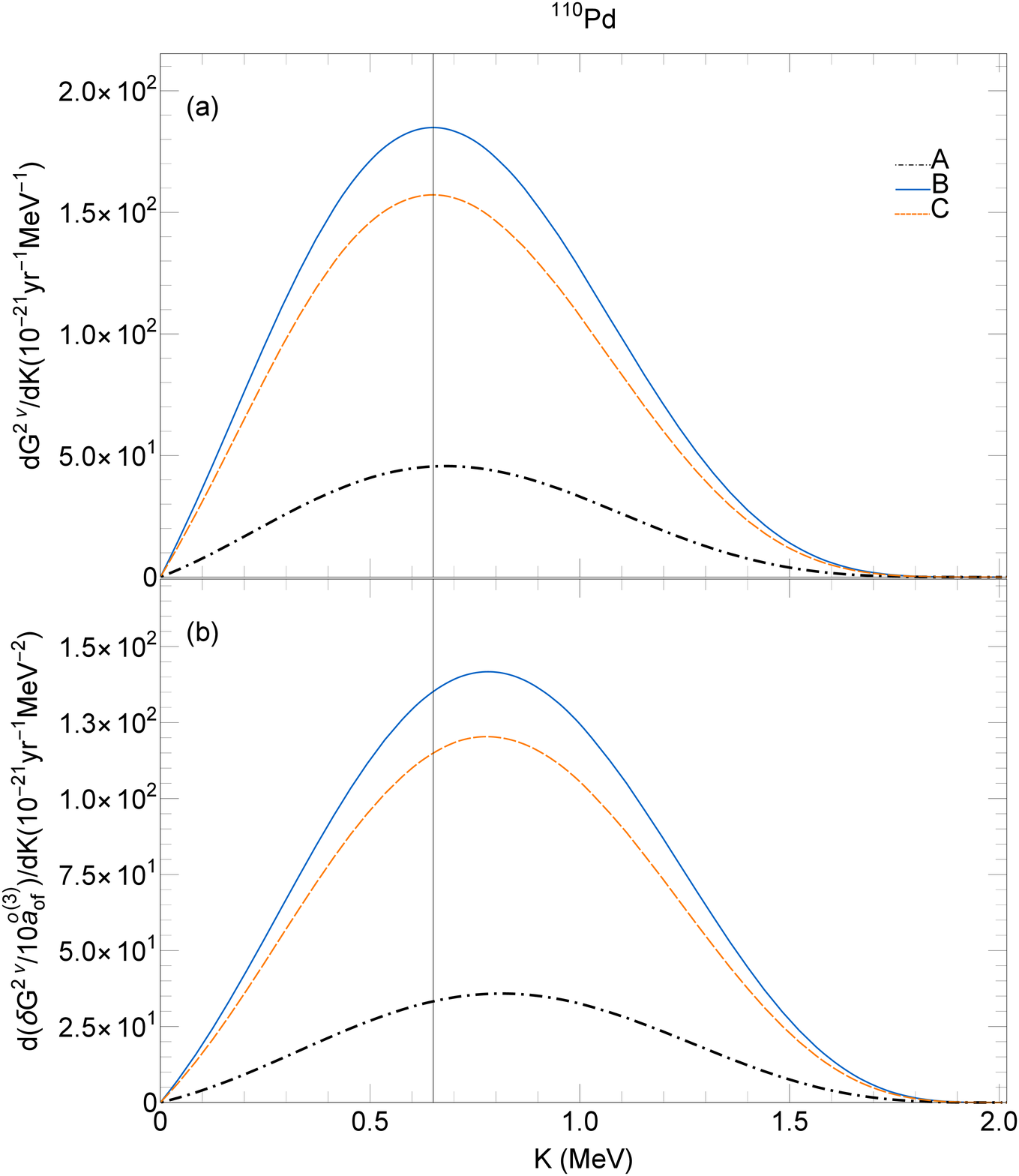}
	\end{subfigure}%
	\begin{subfigure}
		\centering
		\includegraphics[width=0.49\textwidth]{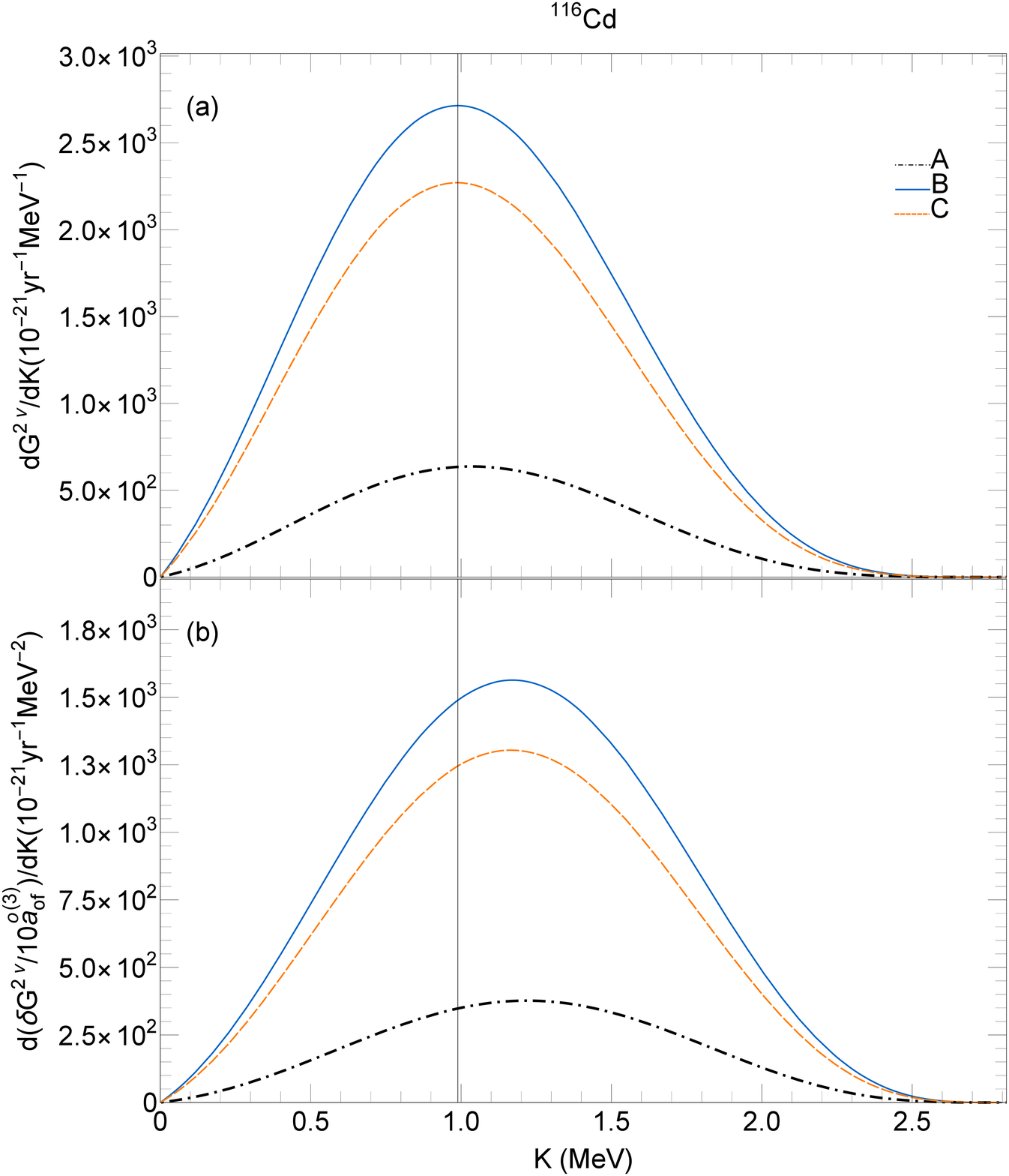}
	\end{subfigure}
	\begin{subfigure}
		\centering
		\includegraphics[width=0.49\textwidth]{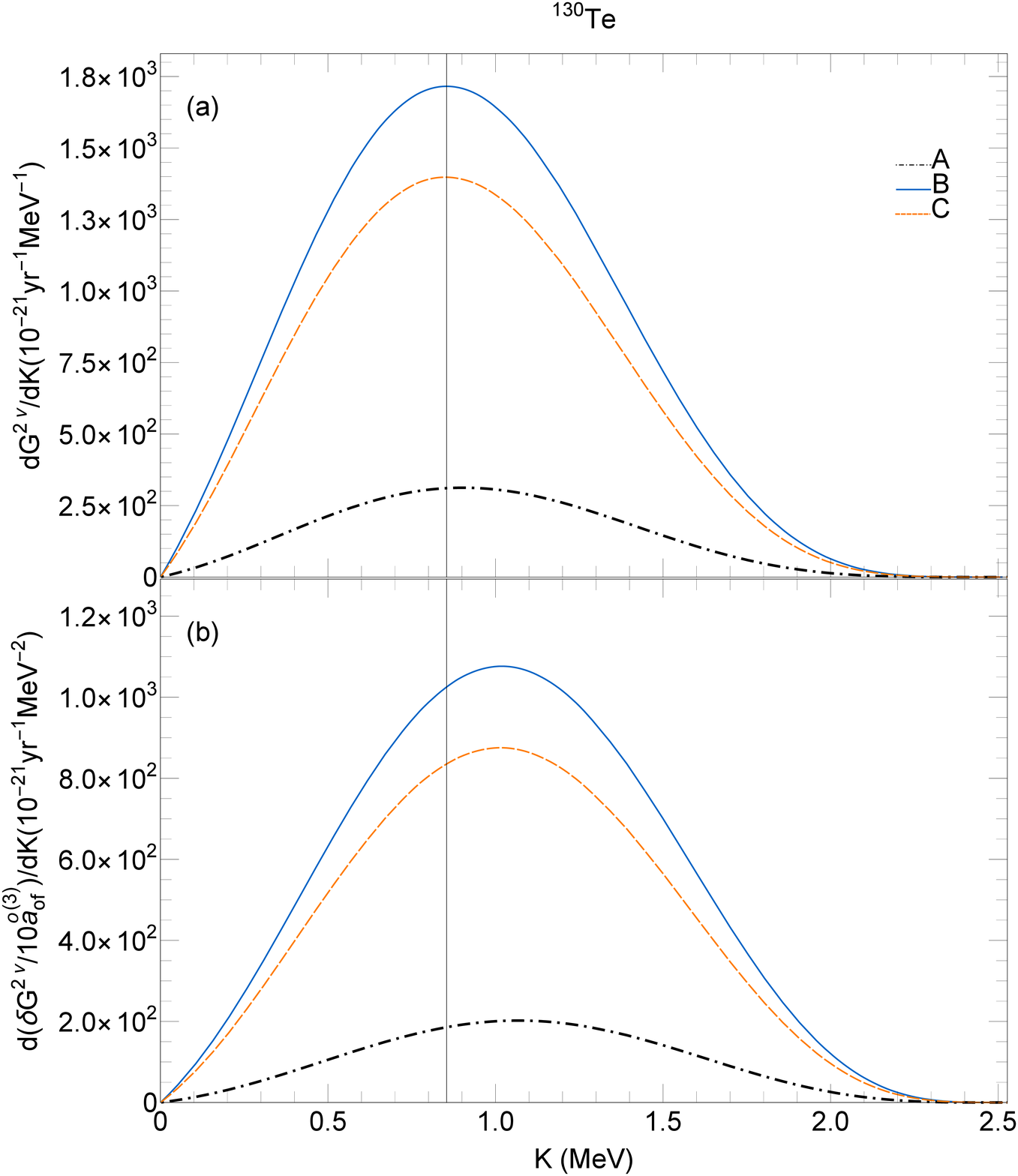}
	\end{subfigure}%
	\begin{subfigure}
		\centering
		\includegraphics[width=0.49\textwidth]{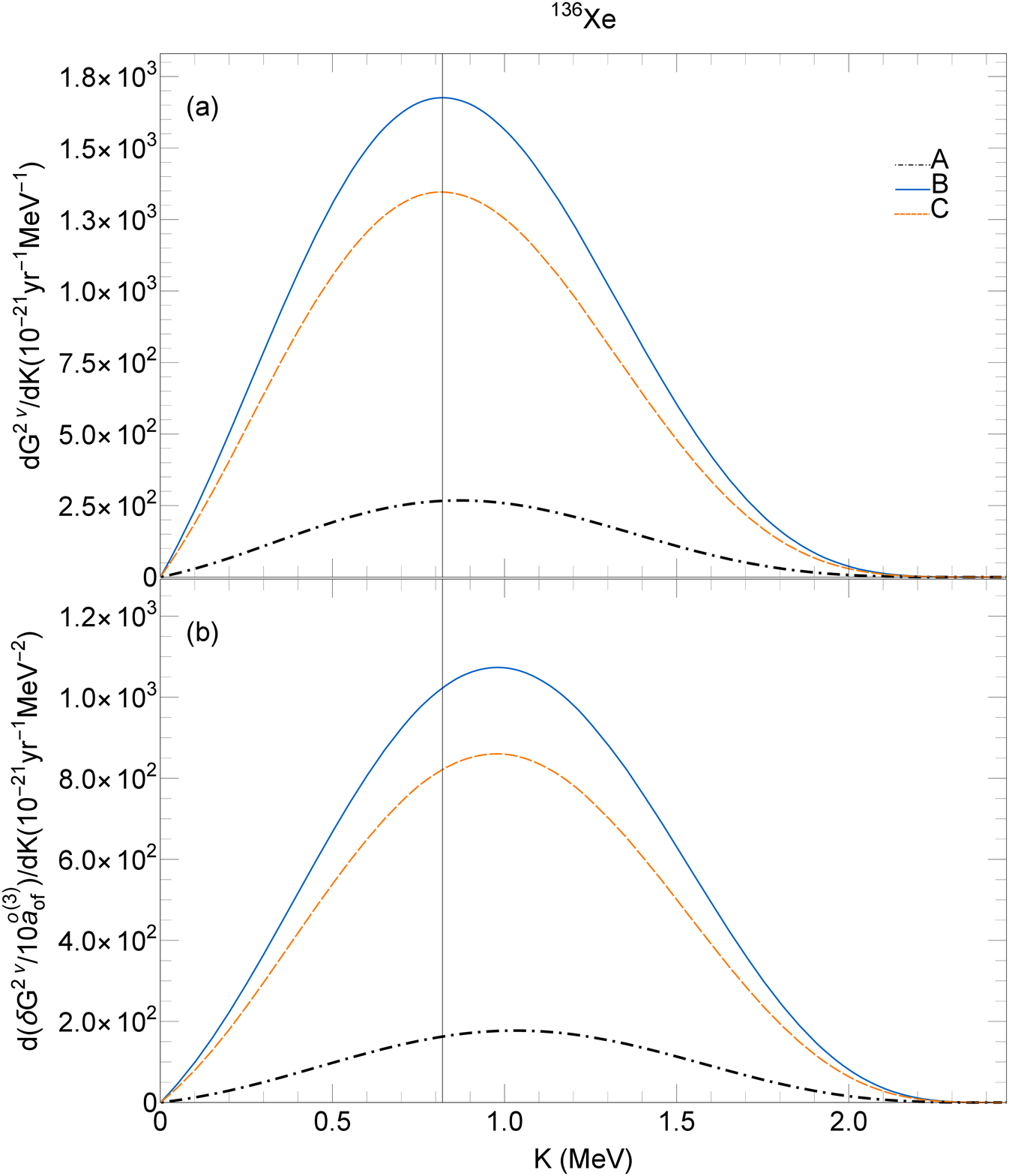}
	\end{subfigure}
	\caption[short]{Summed energy spectra of electrons in the standard $2\nu\beta\beta$ decay (a) and their deviations due to LIV (b) for the nuclei $^{110}$Pd, $^{116}$Cd, $^{130}$Te and $^{136}$Xe.}
	\label{fig:SumSpectra2}
\end{figure}

\section{Electron spectra}
The differences between the PSF values obtained with different Fermi functions are reflected in the electron spectra. To obtain the summed energy spectra of electrons, we changed the integration variables from the individual electrons kinetic energies to their sum $K = \epsilon_1 + \epsilon_{2}$ and their difference $T = \epsilon_1 - \epsilon_2$ with $K\in[0,Q]$ and $T\in[-K, K]$. In \Fref{fig:SumSpectra1} and \Fref{fig:SumSpectra2} are plotted the energy spectra (upper part) and their deviations due to LIV (lower part) corresponding with \Eref{eq:G2nuVS} and \Eref{eq:DeltaG2nuVS}, respectively. The legends in \Fref{fig:SumSpectra1} and \Fref{fig:SumSpectra2} are related to the approximation schemes for the Fermi functions discussed in Section \ref{sec:Fermi}. The spectra corresponding to the LIV component are presented independently of the oscillation-free coefficient $\aof$.   

It can be observed that the use of non-relativistic Fermi functions (A scheme) gives spectra that differ much from those obtained with relativistic Fermi functions (B, C schemes). Also, there are significant differences between the electron spectra obtained with the B and C approximations. Hence, the use of accurate Fermi functions is important for obtaining reliable summed energy spectra of electrons. Further, looking to the lower part of the figures,  one observes that these standard electron spectra should be shifted if LIV effects are present. To highlight this shift, a vertical line passing the maximum of the standard spectrum is present in the figures. We note that, at request, we can provide experimentalists with the numerical data needed to build up theoretical electron spectra.

\section{Quantities of experimental interest}  
In LIV investigations, an important parameter is also the value of the summed energy of electrons where the LIV effect is expected to be maximum. This energy, $K_{m}$, corresponds to the position where the LIV summed energy spectrum of electrons is maximal \cite{Diaz-PRD89}. This quantity is dependent on all the input ingredients taken into account in spectrum computation (Fermi function, Q-value, the inclusion of kinematic factors). As seen in Figures \ref{fig:SumSpectra1} and \ref{fig:SumSpectra2}, the deviation should not be large since all spectra have similar shape. We computed these positions and presented them in \Tref{tab:ExpQuant} together with the ones obtained in \cite{Diaz-PRD89}. The values in the third column were obtained by employing the full treatment of the $2\nu\beta\beta$ decay rate, namely using numerical Fermi functions obtained with approximation scheme C, taking the Q-values given in Appendix and including the kinematic factors.

\begin{table}[!htb]
	\lineup
	\caption{$K_{m}$ - energy where LIV effect is maximal and $\aof G^{2\nu}/\delta G^{2\nu}$ - ratio calculated with formulas (\ref{eq:G2nuVS}) and (\ref{eq:DeltaG2nuVS}).}
	\label{tab:ExpQuant}
	\begin{indented}
		\item[]\begin{tabular}{@{}*{7}{c}}
			\br
			\0\0Nucleus\0\0&\centre{1}{\0\0$K_m$(\rm{keV})\cite{Diaz-PRD89}}\0\0&\centre{1}{\0\0$K_m$(\rm{keV})}\0\0&\centre{1}{\0\0$\aof G^{2\nu}/\delta G^{2\nu}$ ($10^{-6}$\rm{GeV})}\0\0\\
			\mr
			$^{48}$Ca &$1980$&$2002$&$227.199 \pm 0.048$\\
			$^{76}$Ge &$810$&$\0818$&$119.562 \pm 0.009$\\
			$^{82}$Se &$1300$&$1297$&$168.823 \pm 0.195$\\
			$^{100}$Mo&$1320$&$1294$&$172.365 \pm 0.111$\\
			$^{110}$Pd&--&$\0777$&$120.518 \pm 0.420$\\
			$^{116}$Cd&$1200$&$1165$&$162.339 \pm 0.085$\\
			$^{130}$Te&$1050$&$1013$&$148.307 \pm 0.170$\\
			$^{136}$Xe&$1020$&$\0980$&$145.023 \pm 0.269$\\ 			
			\br
		\end{tabular}\\
	\end{indented}
\end{table}

The last column of \Tref{tab:ExpQuant} contains the ratio of the integrals in 
\Eref{eq:G2nuVS} and \Eref{eq:DeltaG2nuVS} which is independent of the LIV coefficient $\aof$ and, as shown
in \cite{CUPID-0-PRD100}, it is the needed input in searching for LIV effects.
The values presented in this column are computed using the same treatment as
for the ones in the third column. We note the difference between our prediction of the ratio $\aof G^{2\nu}/\delta G^{2\nu}$ for $^{82}$Se ($168.823\times 10^{-6}\mathrm{GeV}$) and the one given in \cite{CUPID-0-PRD100} ($213.3\times 10^{-6} \mathrm{GeV}$) to constrain the  $\aof$ coefficient.

\section {Conclusions}
Investigations of LIV effects are currently also conducted in DBD, particularly by searching for deviations of the summed energy spectra of electrons in $2\nu\beta\beta$ decays from their standard form. In the absence of observing such deviations, constraints are placed on the  $\aof$ coefficient that controls the strength of the LIV associated with the time-like component of the countershaded operator in SME theory. For the experimental investigations, theoretical calculations of PSF, summed energy spectra of electrons, and their deviations due to LIV are needed. 
In this work, we provide accurate calculations of these quantities using exact electron wave functions for building the Fermi functions with the inclusion of finite nuclear size and screening effects, Q-values obtained by averaging on values provided by experimental measurements and PSF expressions that include the kinematic factors. Comparing our results with previous ones used in other LIV investigations, we show that the choice of the Fermi functions is the essential ingredient in calculations. We obtained differences up to about $30\%$ even between different relativistic methods of calculation of these functions.  We estimate the uncertainties in the computation of these quantities associated with experimental measurements of the Q-values and with the omission of the kinematic factors in the PSF expressions. Uncertainties in the PSF values due to the use of inaccurate Q-values are nucleus-dependent and are estimated at $8 \times \sigma_Q/Q$. Next, we calculated with our method described in \Sref{sec:ourmethod} and Refs. \cite{SM-2013,MPS-2015} the quantities of experimental interest in LIV analyses, namely the ratios between the standard PSF and their LIV deviations.

Precise calculations can significantly influence the theoretical data needed in LIV analyses. As an example, we found a relevant difference between our obtained value of this ratio and the one used by the CUPID-0 collaboration. Hence, we hope that our theoretical predictions corroborated with a precise analysis of the summed energy spectra of electrons lead to improved constraints on the $\aof$ coefficient that controls the LIV strength of the time-like component of the countershaded operator in SME.

\section{Acknowledgments}
This work has been supported by the grants of the Romanian Ministry of Research and Innovation through the projects UEFISCDI-18PCCDI/2018 and PN19-030102-INCDFM.

\appendix
\section*{Appendix}
\setcounter{section}{1}
One of the most important parameters in the computation of the PSFs is the Q-value of the double-beta decay. In this study, we treated the uncertainty associated with the Q-value as the only source of uncertainty in the PSF. There are multiple experimental values reported in the literature for different double-beta decays, and the use of one or another influences the PSF calculated values and predictions of the electron spectra. The choice of the Q-values used in the PSF calculation is made as follows. For each $2\nu\beta\beta$  decay, we collect the $Q_i$-values from the literature together with their statistical errors ($\delta Q_i$). Then, we calculate an average Q-value and the statistical error for each decay following a procedure presented in \cite{PDG-statistics} that we shortly describe here. 

\begin{table}[!htb]
	\lineup
	\caption{Experimental $Q$ values and the average Q-values for g.s. to g.s. transitions}
	\label{tab:Qvalues}
	\begin{indented}
		\item[]\begin{tabular}{@{}*{7}{l}}
			\br
			Parent&\centre{2}{Measured Energies(keV)}&\centre{2}{Average Value}\\
			\ns
			&\crule{2}&\crule{2}\\
			nucleus&\centre{1}~&\centre{1}~&\centre{1}{$Q\pm\sigma_{Q}$(keV)}&\centre{1}S\\
			\mr
			$^{48}$Ca&$4262.96\0\pm\00.84$\0\cite{Ca-PRC86-2012} &$4268.121\pm0.079$ \cite{Ca-PRC88-2013}&&\\
			&$4267.98\0\pm\00.32$\0\cite{Ca-PRC89-2014}&&$4268.070\pm0.076$&$1.0$\\
			$^{76}$Ge&$2038.56\0\pm\00.32$\0\cite{Ge-PRL67-1991}&$2038.58\0\pm0.31\0$ \cite{Ge-NIMPRA-1993}&&\\
			&$2039.006\pm\00.05$\0\cite{Ge-PRL86-2001} &$2039.04\0\pm0.16\0$ \cite{Ge-PLB662-2008}&&\\
			&$2039.061\pm\00.007$\cite{Ge-PRC81-2010} &&$2039.059\pm0.014$&$2.0$\\
			$^{82}$Se&$2997.9\0\0\pm \00.3$\0\0\cite{Se-PRL110-2013}& &$2997.90\0\pm0.30$&\\
			$^{100}$Mo&$3034.0\0\0\pm\06.0$\0\0\cite{Mo-PhDThesis} &$3034.40\0\pm 0.17\0$ \cite{Ge-PLB662-2008}&$3034.40\0\pm 0.17$&$1.0$\\
			$^{110}$Pd&$2017.85\0\pm\00.64$\0\cite{Pd-PRL108-2012}& &$2017.85\0\pm0.64$&\\
			$^{116}$Cd&$2813.50\0\pm\00.13$\0\cite{Cd-PLB703-2011}&&$2813.50\0\pm 0.13$&\\
			$^{130}$Te&$2527.01\0\pm\00.32$\0\cite{Te-PRC80-2009}& $2527.518\pm0.013$ \cite{Te-PRL102-2009} &&\\
			&$2526.97\0\pm\00.23$\0\cite{Cd-PLB703-2011}&&$2527.51\0\pm0.26$&$2.0$\\
			$^{136}$Xe&$2479.0\0\0\pm79.0$\0\0\cite{Xe-NJP7-2005}&$2462.7\0\0\pm4.3\0\0$ \cite{Xe-IJMS251-2006} &&\\
			&$2457.83\0\pm\00.37$\0\cite{Xe-PRL98-2007}&$2458.73\0\pm0.56\0$ \cite{Xe-PRC82-2010}&$2458.13\0\pm0.41$&$1.3$\\
			
			\br
		\end{tabular}
	\end{indented}
\end{table}

\begin{table}[htb!]
	\lineup
	\caption{\label{tab:PSFVS} Phase space factors (upper part) and their deviations (lower part) computed with numerical Fermi functions and with inclusion of kinematic factors. Relative uncertainties associated with Q-values are displayed in column 3, in terms of $\sigma_{Q}/Q$. Differences introduced by the omission of kinematic factors: $\xi = 
		(G-G^{\tiny\textrm{No VS}})/G$ and $\delta \xi = (\delta G-\delta G^{\tiny\textrm{No VS}})/\delta G$ are also displayed in the last column ($G^{\tiny\textrm{No VS}}$ and $\delta G^{\tiny\textrm{No VS}}$ are the ones from \Tref{tab:PSFAppSchemes}).}
	\begin{indented}
		\item[]\begin{tabular}{@{}*{7}{c}}
			\br
			Nucleus&\centre{1}{$G (10^{-21} {\rm yr}^{-1})$}&\centre{1}{$\sigma_{G}/G$ $\left(\frac{\sigma_{Q}}{Q}\right)$}&\centre{1}{$\xi(\%)$}\\
			\mr
			$^{48}$Ca &$15443.23$&$8.83$&$1.494$\\
			$^{76}$Ge &\0\0\0$48.50$&$8.32$&$0.231$\\
			$^{82}$Se &\0$1604.65$&$8.58$&$0.452$\\
			$^{100}$Mo&\0$3325.32$&$8.53$&$0.382$\\
			$^{110}$Pd&\0\0$138.79$&$8.20$&$0.158$\\
			$^{116}$Cd&\0$2775.55$&$8.43$&$0.286$\\
			$^{130}$Te&\0$1541.74$&$8.32$&$0.208$\\
			$^{136}$Xe&\0$1444.19$&$8.29$&$0.188$\\
			\mr
			\0\0Nucleus\0\0&\centre{1}{\0\0$\delta G/10\mathring{a}_{0f} (10^{-21} {\rm yr}^{-1} {\rm MeV}^{-1})$}\0\0&\centre{1}{\0\0$\sigma_{\delta G}/\delta G$ $\left(\frac{\sigma_{Q}}{Q}\right)$}\0\0&\centre{1}{\0\0$\delta\xi(\%)$}\0\0\\
			\mr
			
			$^{48}$Ca &$6797.21$&$7.94$&$1.800$\\
			$^{76}$Ge &\0\0$40.56$&$7.43$&$0.284$\\
			$^{82}$Se &\0$950.49$&$7.69$&$0.549$\\
			$^{100}$Mo&$1929.23$&$7.64$&$0.465$\\
			$^{110}$Pd&\0$115.16$&$7.31$&$0.195$\\
			$^{116}$Cd&$1709.73$&$7.54$&$0.349$\\
			$^{130}$Te&$1039.56$&$7.43$&$0.255$\\
			$^{136}$Xe&\0$995.83$&$7.39$&$0.231$\\
			
			\br
		\end{tabular}\\
	\end{indented}
\end{table} 
A $Q$-value for a particular DBD is measured in $N$ independent experiments. For the set of measurements obtained, $x_i\pm\delta x_i$,  a Gaussian distribution is considered. Here $x_i=Q_i$ and $\delta x_i=\delta Q_i$ are the value and the error provided by the $i$-experiment, respectively. The weighted average and the corresponding error are calculated with the following equation:
\begin{equation}
\bar{x}\pm\delta \bar{x}=\frac{\sum w_ix_i}{\sum w_i}\pm \left(\sum w_i\right)^{-1/2}
\end{equation} 
where
\begin{equation}
w_i=\frac{1}{(\delta x_i)^2}
\end{equation}
and the sums run over all $N$ experiments.
For each weighted average, we calculated $\chi^2=\sum w_i (\bar{x}-x_i)^2$ and a scale factor, $S$, defined as
\begin{equation}
S=\left(\frac{\chi^2}{N-1}\right)^{1/2}
\end{equation}
to establish if the measurements are indeed from a Gaussian distribution. If the scale factor is less than or equal to $1$, the value of $\delta \bar{x}$ is left unchanged, and the result is accepted. If $S$ is larger than $1$ and the input $\delta x_i$ are all about the same size, then we increase $\delta \bar{x}$ by the scale factor. In the final case of $S$ larger than $1$ and the $\delta x_i$ are of widely varying magnitudes, $S$ is recalculated with only the input for which $\delta x_i\leq3N^{1/2}\delta \bar{x}$. In all cases, the original value of $\bar{x}$ remains unchanged.   

The results of this procedure are displayed in \Tref{tab:Qvalues} where the first column contains the studied nuclei. The second column contains the experimental data available for each nucleus. The average $Q$-values and their uncertainties ($\sigma_{Q}$) obtained with the procedure described above are presented in the third column, while the resulting scaling factor $S$ is shown in the last column.

Since this procedure provides also an averaged uncertainty for each $Q$-value, an estimation of the uncertainties in the PSFs can be easily made as follows. Let $f$ denote both $G$ and $\delta G$ factors which depend on the $Q$-value, which is subject to some uncertainty $\sigma_{Q}$. Then, since $Q$ is the only source of error accounted for in this study, the uncertainty of $f$ can be computed using the formula
\begin{equation}
\sigma_f = \left(\frac{\partial f}{\partial Q}\right)\sigma_{Q}
\end{equation}
As seen in the table, the uncertainty of $Q$ is reflected significantly in the PSF uncertainty.

The use of more accurate expressions for the kinematic factors $\langle K_N\rangle$ and $\langle L_N\rangle$ (\Eref{eq:KnDef} and \Eref{eq:LnDef}, instead of \Eref{eq:KnLnApprox}) also gives differences in the calculations of PSF, summed energy electron spectra and their deviations due to LIV, but with smaller effects than in the case of using different Q-values. In Table \ref{tab:PSFVS} we report the PSF values (upper part) and their deviations (lower part) computed with numerical Fermi functions (approximation C) and inclusion of the kinematic factors.

The third column displays the relative uncertainty of the PSF values in terms of relative uncertainties in the $Q$-values. As a rule of thumb, the relative uncertainty of the PSFS is between 8 and 9 times (Standard) and between 7 and 8 times (LIV) the one of the $Q$-value. The last column of Table \ref{tab:PSFVS} contains the differences in percentages between the calculated values with inclusion or not of the kinematic factors. As seen, there are small deviations associated with this approximation, the largest ones being at $^{48}$Ca nucleus.

\section*{References}
\bibliography{thebibliography}

\end{document}